\documentstyle[preprint,12pt,aps,prd]{revtex}
\begin{document}
\preprint{ISN--93--50}
\draft
\title{SEARCH FOR FLAVOURED MULTIQUARKS\\ IN A SIMPLE BAG MODEL}
\author{S.~Zouzou}
\address{Institut de Physique, Universit\'e de Constantine, Alg\'erie}
\author{J.-M.~Richard}
\address{Institut des Sciences Nucl\'eaires,
 Universit\'e Joseph Fourier - CNRS - IN2P3\\
53, avenue des Martyrs, Grenoble, France}
\date{\today}
\maketitle
\begin{abstract}
We use a bag model to study flavoured mesonic
$(Qq\bar q\bar q)$ and baryonic  $({\overline Q}qqqq)$ states,
where one heavy quark $Q$
is associated with light quarks or antiquarks,
and search for possible stable multiquarks.
No bound state is found. However some states lie not too high above
their dissociation threshold, suggesting the possibility of resonances,
or perhaps bound states in improved models.\end{abstract}
\pacs{ 12.40. Aa, 14.20.Kp, 14.40.Jz}
\section{Introduction}
The search for multiquark states is no longer as fashionable
 as it used to be at the end of the 70's
\cite{ChanBarr,MontRossiVen}, but remains a very
 important issue of the physics of confinement.
A renewed
interest is however observed in the physics of hadrons with heavy flavour,
 and there is a rich literature on heavy-quark symmetry,
 heavy-quark effective theory, etc.\ \cite{Wise,Grinstein}.
 The structure of flavoured
mesons $(Q{\overline Q})$ and baryons $(Qqq)$ and $(QQq)$
 seems  better understood now, and reliable extrapolations towards
 exotic configurations like the ``Tetraquark''$(Qq\bar q\bar q)$
 and the ``Pentaquark''  $({\overline Q}qqqq)$
 sectors will become feasible.
(Here and throughout this paper, $Q$ denotes a charmed or bottom quark,
 $q$ a light quark, which is either the strange $s$,
  or an ordinary $q_0=u$ or $d$.)
 It is  useful to summarize what can be learned on these flavoured
 exotics from simple models before considering more sophisticated
 approaches.

Many papers have been devoted to multiquark spectroscopy
in simple potential models or in bag models of various types.
It is impossible to list all relevant references.
Let us mention, for instance: in the ``Hexaquark'' or dibaryon sector,
the pioneering work by Jaffe on the H particle $(ssuudd)$ \cite{JaffeH},
and subsequent studies by the Nijmegen group \cite{Mulders,Aerts}; in the
5-quark sector, again the work of the Nijmegen group \cite{Mulders,Aerts}
or, more recently by Lipkin and the Grenoble group on the flavoured
pentaquark \cite{Pentaquark}; in the 4-quark sector, the numerous
speculations about scalar mesons \cite{JaffeMeson} or the observation that
flavour-independence inevitably leads to stable tetraquark states $(QQ\bar
q\bar q)$ if the mass ratio $m(Q)/m(q)$ is large enough \cite{TetraQuark}.

The various tentative multiquarks experience different aspects
 of quark forces. A deeply bound state would test how the
Coulomb-plus-linear potential of quarkonium applies when more than two or
three constituents are lying close together. On the other hand, the loosely
bound meson--meson molecules, predicted by several authors
\cite{Dooley,BarnesMol,Tornqvist,Weinstein,Manohar}, would be more sensitive
to the long-range part, or say, the Van-der-Waals or Yukawa regime of the
interaction.

The analogy with atomic physics might provide some guidance,
 to guess which flavour configurations are the most favourable for stable
multiquarks. The reason is that in both cases, we have a central potential
that does not depend on the mass of the constituents, a property called
universality, or flavour independence. For instance, the H$^-$ ion
$(pe^-e^-)$ with $m(p)\gg m(e)$ has a slightly larger relative energy
 (as compared to its threshold) than the positronium ion $(e^+e^-e^-)$
\cite{ArmourB}; in
the 4-charge sector, the so-called positronium hybride  $(pe^+e^-e^-)$ is
stable against dissociation into $(pe^-)+(e^+e^-)$, by a small margin
\cite{Hybride}: these
observations are felt as an encouragement to study singly-flavoured
multiquarks in hadron spectroscopy.

Of course, there are dramatic differences between colour forces
 and Coulomb forces. In particular, the spin-dependent corrections play a
more important role in hadrons than in atoms. This led to speculations on
states bound essentially by the coherent effect of hyperfine forces (also
called chromomagnetic forces). A systematic study of chromomagnetism for
multiquarks was carried out by de Swart et al.\ \cite{Mulders,Aerts}, and
further developed by several authors
\cite{LichtMulti,SemayBSB,HogaasenBologna}. We shall use their techniques
and results along this work.

To summarize, collective binding of multiquark configurations
might result either from peculiar asymmetries of the constituent masses,
 as in molecular physics, or from a favourable arrangement of the spins.
When building a model, one should combine the effects of electric
confinement and magnetic forces. In the potential model approach, the former
is described by a flavour-independent central potential, and the latter by
the spin--spin potential. In the bag picture we shall adopt, we have quarks
moving inside a cavity, and chromoelectric as well as chromomagnetic
interactions between them.

Our paper is organized as follows. In Sec.~\ref{Model}, we briefly
present the bag model we shall use.
In Sec.~\ref{Tetraquark}, we apply this model
 to tetraquarks with a single heavy flavour, $(Qq\bar q\bar q)$, while the
pentaquark case $(\overline{Q}qqqq)$ is presented in
Sec.~\ref{sec-Pentaquark}. The discussion in Sec.~\ref{Conclusions} is
guided by the comparison with potential models. We conclude that our model
does not predict the existence of stable flavoured multiquarks, but can
select the most favourable configurations,
tyo be studied in a more elaborated theoretical framework.

\section{Model}
\label{Model}
The bag model is well known, and hardly needs to be explained in detail. We
shall restrict ourselves to a quick review, with however all the
necessary equations, to make this paper self contained.
For a review on the bag model, see, e.g.\
Refs.~\cite{JaffeRev,DeTarRev,ThomasRev}.

\subsection{The MIT bag}
\label{MIT-bag}
The original fit by the MIT group
\cite{DeGrand} of groundstate mesons and baryons with $u$, $d$ and
$s$ quarks corresponds to the static cavity approximation of the bag model
\cite{Chodos}.
The quark wave functions are given by the free Dirac equation inside a
sphere of radius $R$. The linear boundary conditions fix the wave number
$x/R$ of each quark species as the lowest solution of
\begin{equation}
\label{x-equation}
x=\left(1-mR-\sqrt{x^2+m^2R^2}\right)\tan x.
\end{equation}
The corresponding energy is $\omega=(x^2+m^2R^2)^{1/2}/R$.
For a given radius, the hadron energy, \begin{equation} \label{MIT-energy}
E(R)={4\over3}\pi BR^3+\sum_i\omega_i-{Z_0\over R}+\delta E_{\rm e}+\delta
E_{\rm m}, \end{equation} combines the volume energy, the kinetic energy, the
zero-point energy (which presumably includes many other corrections in an
effective way), and (chromo-)electric and magnetic corrections. The electric
term is
\begin{equation}
\label{MIT-electric-1}
\delta E_{\rm e}={\alpha_{\rm c}\over 2R}\sum_{i,j}
\tilde{\lambda}_i\cdot\tilde{\lambda}_jf_{i,j}.
\end{equation}
It vanishes if a single quark species is involved, since the colour operators
$\tilde{\lambda}_i\cdot\tilde{\lambda}_j$ sum up to zero for a colour singlet.
For $K$, $K^*$, $\Lambda$, $\Sigma$,
$\Sigma^*$, $\Xi$, or $\Xi^*$, the value of
\begin{equation}
\label{MIT-electric-2}
\delta E_{\rm e}={8\alpha_{\rm c}\over 3R}
\left(f_{s,s}+f_{0,0}-2f_{0,s}\right)
\end{equation}
turns out to be very small. The strength $f_{i,j}$ is calculated out of the
quark densities
\begin{mathletters}
\label{MIT-electric-3}
\begin{eqnarray}
 f_{i,j}&=&\int_0^1\rho_i\rho_j{{\rm d}u\over u^2}\\
\rho&=&{\omega(x'-\sin^2x'/x')-m(\cos x'\sin x'-\sin^2x'/x')\over
\omega(x\phantom{'}-\sin^2x\phantom{'}/x\phantom{'})-m(\cos x\phantom{'}\sin
x\phantom{'}-\sin^2x\phantom{'}/x\phantom{'})},
\end{eqnarray}
\end{mathletters}
where $x'=ux$. The magnetic term reads
\begin{equation}
\label{MIT-magnetic-1}
 \delta E_{\rm m}=-{3\alpha_{\rm c}\over R}\sum_{i<j}
\vec\sigma_i\cdot\vec\sigma_j\tilde{\lambda}_i\cdot\tilde{\lambda}_jg_{i,j}.
\end{equation}
with the strength $g_{i,j}$ expressed in terms of (reduced) magnetic moments
and densities
\begin{mathletters}
\label{MIT-magnetic-2}
\begin{eqnarray}
g_{i,j}&=&\tilde{\mu}_i\tilde{\mu}_j+2\int_0^1\bar{\mu}_i
\bar{\mu}_j{\rm d}u,\\
 \tilde{\mu}&=&{4\omega R+2mR-3\over12\omega R(\omega R-1)+6mR},\\
\bar{\mu}&=&{(\omega R-mR)x\over\left[
2\omega R(\omega R-1)+mR\right]\sin^2x}\left(
{1-3\sin(2x')/4x'+\cos(2x')/2\over3x'}\right),
\end{eqnarray}
\end{mathletters}
where, again, $x'=ux$.
As a result of the non-linear boundary conditions,
the bag energy (\ref{MIT-energy}) is minimized with respect to $R$. The
results given in Table III of Ref.~\cite{DeGrand} correspond to the set of
parameters and masses
\begin{mathletters}
\label{MIT-set}
\begin{eqnarray}
 B^{1/4}&=&0.145\,{\rm GeV},\quad Z_0=1.81,\quad\alpha_{\rm c}=0.55,\\
m_0&=&0,\qquad m_s=0.279\,{\rm GeV},
\end{eqnarray}
\end{mathletters}

\subsection{Bag with one heavy quark}
\label{Bag-Heavy-Light}
Numerous improved or modified versions of the MIT bag model
 have been developped, with a variety of purposes. One problem was to adapt
the model to include heavy quarks. In the case of quarkonium states
$(Q\overline Q)$, or triple-flavour baryons $(QQQ)$, an ``adiabatic'' bag
model was proposed \cite{ReportsHK,AdiabHH,AdiabMes,AdiabBar,Baacke}: for
fixed interquark separations, the shape and size of the bag is adjusted to
minimize the energy of the gluon field, and this minimum is read as the
interquark potential, and inserted into the Schr\"odinger equation.

The case of hadrons with one heavy   quark was
 treated by Izatt et al.\ \cite{Izatt}. They first introduce a running
coupling constant, which depends on the bag radius (we keep here the MIT
notation $\alpha_{\rm c}= \alpha_{\rm s}/4$)
\begin{equation}
\label{alpha-Izatt}
\alpha_{\rm c}={\pi\over18\ln\left(1+1/(\Lambda R)\right)}.
\end{equation}
Another change is that the bag radius $R$ is adjusted in
the approximation where only the volume, kinetic (and mass), and zero-point
energies are included. The electric and magnetic terms are then computed
with this radius already fixed.

The model holds for ordinary hadrons, but in this case a
centre-of-mass correction is applied to the total energy $E$,
leading to the hadron mass
\begin{equation}
\label{com-Izatt}
E'=\left[E^2-\sum_i{\bar x_i^2\over R^2}\right]^{1/2}.
\end{equation}

For hadrons with charm or beauty, it is assumed that the heavy quark
stays at the centre of the bag. No centre-of-mass correction is applied. The
electric and magnetic terms involving the heavy quark are modified as
follows ($M$ denotes the mass of the heavy quark, $m$ that of a light quark)
\begin{mathletters}
\label{elec-magn-heavy}
\begin{eqnarray}
f_{M,M}&=&-1\\
f_{M,m}&=&Nx\omega R\left[{\sin^2x\over2x^2}
+{\sin x\cos x\over x}-{3\over2}+\int\nolimits_0^{2x}{1-\cos u\over u}{\rm
d}u\right]-{NmR} \left[{\sin^2x\over2x}-{x\over2}\right]\\
N^{-1}&=&\omega R(x-\sin^2x/x)-mR\sin x(\cos x-\sin x/x)\\
\tilde \mu(M)&=&1/(2MR)\\
g_{M,m}&=&\tilde \mu(M)\tilde \mu(m)\left[1+{2x^3/3-x+\sin(2x)/2\over
x-3\sin(2x)/4+x\cos(2x)/2}\right]
\end{eqnarray}
\end{mathletters}

We shall use the model of Ref.~\cite{Izatt} with a slightly modified
set of parameters
\begin{mathletters}
\label{Izatt-set}
\begin{eqnarray}
B^{1/4}&=&0.1383\,{\rm GeV},\quad\Lambda=0.400\,{\rm GeV},\quad Z_0=0.574,\\
\bar x_0&=&x_0\ (=2.042..),\qquad\bar x_s=2.3,\\
m_0&=&0\quad m_s=0.273\,{\rm GeV},\quad m_c=2.004\,{\rm GeV},\quad
m_b=5.360\,{\rm GeV}.
\end{eqnarray}
\end{mathletters}

A reasonable description of the spectrum of ordinary, charmed or
beautiful hadrons can be achieved. This is illustrated in Table \ref{Tab0}.
We are not too worried anyhow about obtaining a very good fit, since we
shall use our computed masses instead of the experimental ones, when
estimating the thresholds of multiquark candidates. One can check in Table
\ref{Tab0} that the model does not overestimate the strength of magnetic
forces, which are a potential source of collective binding.

This simple model explicitly incorporates the property
of ``heavy quark symmetry'', since the bag radius and the light-quark wave
function are exactly the same for $(c\bar q)$ and $(b\bar q)$, or $(cqq)$
and $(bqq)$. We do expect some recoil corrections in the case of charm, but
we hope they are somehow incorporated in fixing the values of
the parameters.

We note an appreciable freedom in fixing the values of the parameters,
when comparing the MIT values (\ref{MIT-set}) with our set of
parameters(\ref{Izatt-set}). The bag model does not constrain
$\alpha_{\rm c}$, $B$, and $Z_0$ too much when one tries to reproduce the
observed masses, and introducing explicit centre-of-mass corrections implies
some new tuning. A similar discrepancy is observed when comparing
(\ref{Izatt-set}) to the parameters used in Ref.~\cite{AdiabMes} for
charmonium: $B^{1/4}=0.235\,$GeV, and $m_c=1.35\,$GeV.

\section{Tetraquark}\label{Tetraquark}
The  model of Sec.\ \ref{Bag-Heavy-Light} can be applied to
$(Qq\bar q\bar q)$ configurations.  We look at the $J^P=0^+$ groundstate.
The main changes, as compared to ordinary mesons and baryons,  concern the
electric and magnetic terms, which are now operators in the space of the
possible spin--colour wave functions. One has to diagonalize the cumulated
contribution of these electric and magnetic corrections. Details are
provided in Appendix.

An example is examined in Table \ref{Tab1}. The various contributions to
the energies are dispalyed for a $(cs\bar u\bar u)$ state, and for the
hadrons constituting its threshold. The bag radius is larger for $(cs\bar
u\bar u)$ than for $(c\bar u)$ and $(s\bar u)$ mesons, and this pushes
$(cs\bar u\bar u)$ above its dissociation threshold.

The results for all flavour configurations are shown in Table \ref{Tab2}.
The best candidate seems $(Qs \bar u\bar d)$ with isospin $\bar I=0$. It
benefits from the electric interaction between the heaviest quarks, $Q$ and
$s$, and from the magnetic attraction between the lightest, $u$ and $d$,
whereas its threshold $(Q\bar q_0)+(s\bar q_0)$ combines the constituents in
a less favourable way. However $(Qs \bar u\bar d)$ is not bound in our bag
model.

\section{Pentaquark}\label{sec-Pentaquark}
The calculation of $({\overline Q} qqqq)$ can be done with
the same bag model. Details on the electric and magnetic terms, and on the
spin-colour wave functions are given in Appendix.

{}From previous studies \cite{Pentaquark}, we know that the
most favourable configuration  are these where the quarks $(qqqq)$ are in a
triplet state of the flavour group SU(3)$_{\rm F}$, and essentially in a
state of spin and parity  $j^p=0^+$. We thus restrict ourselves to study the
flavour configurations $({\overline Q} udss)$ and $({\overline Q}sudd)$ (and
its charge symmetric with $u\leftrightarrow d$) with total spin and parity
$J^P=(1/2)^-$.

The results are displayed in Table \ref{Tab3}. We essentially
agree with Ref.\ \cite{Flecketal}, and our conclusions are identical: the
pentaquark is bound in the limit where $m(Q)=\infty$ and $m(s)=m(q_0)$, but
the stability does not survive the heavy quark mass being finite and
in the first place flavour SU(3)$_{\rm F}$ symmetry being broken.

\section{Comments and conclusions}
\label{Conclusions}
The multiquark masses listed in  Tables~\ref{Tab2}--\ref{Tab3} result from
the whole model. It seems useful to analyze  the role of the
various contributions.

Let us first consider  the crudest version of the bag model, with two terms
in the energy, corresponding to the volume and quark contributions
\begin{equation}
\label{crude-1}
E=bR^3+aR^{-1},
\end{equation}
where $b=4\pi B/3$. If $a$ is independent of the radius $R$,
the minimization with respect to  $R$ leads to a kind of virial theorem,
where the  bag energy is four times larger than the volume term, and
exhibits a very simple behaviour \begin{equation}
\label{crude-2}
E_{\text{bag}}=\min_{R}\left(E\right)\propto a^{3/4}.
\end{equation}
Thus, in a model with $n$ massless quarks in a cavity and without correction
for zero-point energy, $a=nx_0$ grows as $n$, and the hadron mass behaves as
$M(n)\propto n^{3/4}$. This would mean $M(4)<2M(2)$, $M(6)<2M(3)$, etc.,
i.e.\ stability for many multiquarks.

In actual models, $a$ depends slightly on $R$ for quarks of finite mass, and
the term $a/R$ incorporates the electric and magnetic corrections, as well as
the zero-point energy.
The $n$ dependence is thus more involved.

Already, when going from $n=2$ (mesons) to $n=3$ (baryons), the MIT model or
its modified versions do not behave as $M\propto n^{3/4}$, which would give
\begin{equation}
\label{crude-3}
{M(qqq)\over3}<{M(q\bar q)\over2},
\end{equation}
in contradiction with simple potential models \cite{JMRrep}, and
with experimental data \cite{PDG} (compare for instance the spin-averaged
mass of $\pi$ and $\rho$ with that of $N$ and $\Delta$, or $\Phi(1020)$ with
$\Omega(1672)$). In other words, fitting of mesons and
baryons  simultaneously requires an empirical $Z_0/R$ correction
\cite{JaffeRev}, and this prevents  a proliferation of multiquarks in
the bag model.

An illustration is provided in Table \ref{Tab4},
where the tetraquark calculation is repeated with the zero-point energy
constant $Z_0$ set to zero. We no longer fit  the hadron masses, but
this is not too important, since stability is estimated using the calculated
masses for the ordinary hadrons. We observe in Table \ref{Tab4} that the
tetraquark mass becomes closer to the corresponding threshold, and sometimes
even smaller than the threshold.

Another crucial ingredient for binding multiquarks is  the strength
of 2-body  correlations at short distances. In investigatory scans of the
spin-flavour space, one sometimes considers simplified Hamiltonians of the
type
\begin{equation}
\label{NaiveChromo}
H=K\sum_{i<j}\vec\sigma_i\cdot\vec\sigma_j\,
\tilde\lambda_i\cdot\tilde\lambda_j\,a_{i,j}, \end{equation}
and looks at the spectrum obtained with some crude ansatz
for $a_{i,j}$. For instance $a_{i,j}$ can be taken as independent
of the flavour of the quarks $i$ and $j$, and of the particular hadron one
considers; or one can assume $a_{i,j}=a'/(m_im_j)$, and take $a'$ as being
constant and universal; etc.

Specific dynamical calculations of the correlation factors $a_{i,j}$
have been performed, or are implicit in some
of the previous works on multiquarks. Most authors used
non-relativistic potential models, with a simple prescription
to extrapolate the quark--antiquark potential of mesons
toward the multiquark sector, for instance
\begin{equation}
\label{lambda-lambda}
V=-{3\over16}\sum_{i<j}\,\tilde\lambda_i\cdot\tilde\lambda_j\,
V_{Q\overline Q}(r_{ij}),
\end{equation}
in naive analogy with atomic physics.

An interesting result of such calculations is that the short-range 2-body
correlation is generally smaller in a multiquark state than in ordinary
hadrons. This tends to weaken or to kill the binding of multiquark predicted
by simple chromomagnetic models like (\ref{NaiveChromo}).

In our bag model, we observe a similar effect:
the bag radius $R$ is larger for  multiquarks than for ordinary hadrons, and
this lowers the strength of chromomagnetic terms.

To summarize, we confirm the conclusions of previous studies
of the flavoured pentaquark: the stability predicted on the basis of simple
chromomagnetic calculations disappears due to the lack of short-range
correlations between light quarks.
However, the configuration $(\bar b uuds)$ is found not too far from being
bound, and deserves to be further studied.
Similarly, none of the  possible tetraquark configurations $(Qq\bar
q\bar q)$ is found stable in our model, but those where the antiquarks have
strangeness $S=0$ and isospin $\bar I=0$ seem the best candidates in the
event that improving the model would provides some additional attraction.

\acknowledgments
We would like to thank our colleagues of Grenoble for
several fruitlful discussions and for the hospitality extended to S.R.Z.\
during a visit where this work was completed, and
R.~Barrett for useful comments on the manuscript. J.M.R.
benefitted from the warm hospitality of the ECT* at Trento.

\appendix
\section{}
We now provide some details about the spin--colour wave functions, and the
electric and magnetic terms, for tetraquark and pentaquark states.
The simplest case is $(Qs\bar s\bar s)$ or $(Qq_0\bar q_0\bar q_0)$, with the
antidiquark in an isospin $\bar I=1$ state. We can use the spin--colour basis
\begin{equation}
\label{Tetra-spin-colour-1}
\vert1\rangle=\vert(\bar3,3),(1,1)\rangle,\qquad\vert2\rangle=\vert(6,\bar6),(0,0)\rangle,
\end{equation}
to combine the diquark, of colour $\bar 3$ or $6$ and spin 0 or 1, with the
antidiquark, whose spin and colour are correlated by the Pauli principle. The
electric term is scalar, with value
\begin{equation}
\label{Tetra-electric-1}
\delta E_{\rm e}=-{8\alpha_{\rm
c}\over3R}\left[2f_{M,m}-f_{m,m}-f_{M,M}\right],
\end{equation}
and one has to diagonalize the colour--spin operator arising in the magnetic
term
\begin{equation}
\label{Tetra-magnetic-1}
\delta E_{\rm m}=-{3\alpha_{\rm
c}\over2R}\left(g_{m,m}+g_{M,m}\right)\pmatrix{16/3&8\sqrt6\cr8\sqrt6&-8}
\end{equation}
The eigenvalue of interest in the above matrix is
$\left(-8 + 8\,\sqrt{241}\right)/ 6\simeq19.4$

Another simple case is $(Qq_0\bar q_0\bar q_0)$ with isospin $\bar I=0$ for the
antiquarks. The basis is now
\begin{equation}
\label{Tetra-spin-colour-2}
\vert3\rangle=\vert(\bar3,3),(0,0)\rangle,\qquad
\vert4\rangle=\vert(6,\bar6),(1,1)\rangle.
\end{equation}
The electric term is unchanged. The magnetic term is similar to
(\ref{Tetra-magnetic-1}), besides the matrix which is now
\begin{equation}
\label{Tetra-magnetic-2}
\pmatrix{16&8\sqrt6\cr 8\sqrt6&88/3},
\end{equation}
and whose largest eigenvalue is $\left(136 + 8\,\sqrt{241}\right)/6\simeq43.4$.

For $(Qq_0\bar s\bar s)$, or $(Qs\bar q_0\bar q_0)$ with $\bar I=1$, the basis
(\ref{Tetra-spin-colour-1}) is still appropriate, and the electric term
diagonal. The matrix elements of interest are
\begin{eqnarray}
\label{Pqss}
\left(\delta E_{\rm e}\right)_{1,1}&=&
{8\alpha_{\rm c}\over 3R}\left(f_{1,1}+f_{2,2}+f_{3,3}-f_{1,2}-f_{1,3}-f_{2,3}
\right),\nonumber\\
\left(\delta E_{\rm e}\right)_{2,2}&=&
{4\alpha_{\rm c}\over
3R}\left(2f_{1,1}+2f_{2,2}+5f_{3,3}+f_{1,2}-5f_{1,3}-5f_{2,3}
\right),\nonumber\\
\left(\delta E_{\rm m}\right)_{1,1}&=&
-{8\alpha_{\rm c}\over R}\left(2g_{1,3}-g_{1,2}+2g_{2,3}-g_{3,3}\right),\\
\left(\delta E_{\rm m}\right)_{2,2}&=&
 {12\alpha_{\rm c}\over R}\left(g_{1,2}+g_{3,3}\right),\nonumber\\
\left(\delta E_{\rm m}\right)_{1,2}&=&
-{12\sqrt6\,\alpha_{\rm c}\over R}\left(g_{1,3}+g_{2,3}\right),\nonumber
\end{eqnarray}
where 1 stands for the heavy quark, 2 for $q_0$ (for $s$), and 3 for $\bar s$
(for $\bar q_0$), in the case
of $(Qq_0\bar s\bar s)$ (respectively $(Qs\bar q_0\bar q_0)$ with $\bar I=1$).

For  $(Qs\bar q_0\bar q_0)$ with $\bar I=0$, we use the basis
(\ref{Tetra-spin-colour-2}) and the electric and magnetic terms of interest are
\begin{eqnarray}
\label{Psqq}
\left(\delta E_{\rm e}\right)_{3,3}&=&
{8\alpha_{\rm c}\over 3R}\left(f_{1,1}+f_{2,2}+f_{3,3}-f_{1,2}-f_{1,3}-f_{2,3}
\right),\nonumber\\
\left(\delta E_{\rm e}\right)_{4,4}&=&
{4\alpha_{\rm c}\over
3R}\left(2f_{1,1}+2f_{2,2}+5f_{3,3}+f_{1,2}-5f_{1,3}-5f_{2,3}
\right),\nonumber\\
\left(\delta E_{\rm m}\right)_{3,3}&=&
-{24\alpha_{\rm c}\over R}\left(g_{3,3}+g_{1,2}\right),\\
\left(\delta E_{\rm m}\right)_{4,4}&=&
- {4\alpha_{\rm c}\over
R}\left(g_{1,2}+10g_{1,3}+10g_{2,3}+g_{3,3}\right),\nonumber\\
\left(\delta E_{\rm m}\right)_{3,4}&=&
-{12\sqrt6\,\alpha_{\rm c}\over R}\left(g_{1,3}+g_{2,3}\right),\nonumber
\end{eqnarray}
where we use the labeling $1=Q$, $2=s$, and $3=q_0$.

For $(Qq_0\bar s\bar q_0)$ or $(Qs\bar q_0\bar s)$ (the labeling corresponds to
this order, and is such that $m_2=m_4$), the whole basis
(\ref{Tetra-spin-colour-1},\ref{Tetra-spin-colour-2}) contributes, and the
non-vanishing matrix elements are
\begin{eqnarray}
\label{Pqsqe}
\left(\delta E_{\rm e}\right)_{1,1}&=&
\left(\delta E_{\rm e}\right)_{3,3}=
{4\alpha_{\rm c}\over3R}
\left(2f_{1,1}+3f_{2,2}-3f_{1,2}-f_{1,3}-3f_{2,3}+2f_{3,3}\right)\nonumber\\
\left(\delta E_{\rm e}\right)_{1,4}&=&
\left(\delta E_{\rm e}\right)_{2,3}=
{2\sqrt2\,\alpha_{\rm c}\over R}\left(f_{1,2}-f_{1,3}-f_{2,2}+f_{2,3}\right)\\
\left(\delta E_{\rm e}\right)_{2,2}&=&
\left(\delta E_{\rm e}\right)_{4,4}=
{2\alpha_{\rm c}\over3R}
\left(4f_{1,1}-3f_{1,2}-5f_{1,3}+3f_{2,2}-3f_{2,3}+4f_{3,3}\right)\nonumber
\end{eqnarray}
and
\begin{eqnarray}
\label{Pqsqm}
\left(\delta E_{\rm m}\right)_{1,1}&=&
-{8\alpha_{\rm c}\over R}
\left(g_{1,3}+g_{2,2}\right)\nonumber\\
\left(\delta E_{\rm m}\right)_{1,2}&=&
-{6\sqrt6\,\alpha_{\rm c}\over R}
\left(g_{1,2}+g_{1,3}+g_{2,2}+g_{2,3}\right)\nonumber\\
\left(\delta E_{\rm m}\right)_{1,3}&=&
{4\sqrt3\,\alpha_{\rm c}\over R}
\left(g_{1,2}-g_{1,3}-g_{2,2}+g_{2,3}\right)\nonumber\\
\left(\delta E_{\rm m}\right)_{1,4}&=&
{12\sqrt2\,\alpha_{\rm c}\over R}
\left(g_{1,2}-g_{1,3}-g_{2,2}+g_{2,3}\right)\nonumber\\
\left(\delta E_{\rm m}\right)_{2,2}&=&
{12\alpha_{\rm c}\over R}
\left(g_{1,2}+g_{2,3}\right)\\
\left(\delta E_{\rm m}\right)_{2,4}&=&
{10\sqrt3\,\alpha_{\rm c}\over R}
\left(g_{1,2}-g_{1,3}-g_{2,2}+g_{2,3}\right)\nonumber\\
\left(\delta E_{\rm m}\right)_{3,3}&=&
-{24\alpha_{\rm c}\over R}
\left(g_{1,2}+g_{2,3}\right)\nonumber\\
\left(\delta E_{\rm m}\right)_{3,4}&=&
-{6\sqrt6\,\alpha_{\rm c}\over R}
\left(g_{1,2}+g_{1,3}+g_{2,2}+g_{2,3}\right)\nonumber\\
\left(\delta E_{\rm m}\right)_{4,4}&=&
-{4\alpha_{\rm c}\over R}
\left(6g_{1,2}+5g_{1,3}+5g_{2,2}+6g_{2,3}\right)\nonumber
\end{eqnarray}

For the Pentaquark $({\overline Q} qqqq)$, the quark pairs $(2,3)$ and (4,5)
are either in colour $\bar 3$ or $6$, provided the four quarks form a colour 3
state, which neutralizes the colour of the heavy antiquark. The spins of these
pairs are either 0 or 1,  the spin of $(qqqq)$ being mostly $j=0$, with a small
$j=1$ admixture, so that the spin of the pentaquark is $J=1/2$. We consider the
configurations $({\overline Q}ssud)$ and $({\overline Q} uuds)$ (and
$({\overline Q}ddus)$, its isopsin partner) for which the quarks 2 and 3 are
identical  and thus have their colours and spins correlated.
A possible  basis is
\begin{equation}
\matrix{
|1\rangle=\left|\left(\bar3,\bar3\right);(1,1)0\right\rangle\quad&|5\rangle=\left|\left(\bar3,6\right);(1,1)1\right\rangle\cr
|2\rangle=\left|\left(\bar3,6\right);(1,1)0\right\rangle\quad&|6\rangle=\left|\left(6,\bar3\right);(0,1)1\right\rangle\cr
|3\rangle=\left|\left(6,\bar3\right);(0,0)0\right\rangle\quad&|7\rangle=\left|\left(\bar3,\bar3\right);(1,0)1\right\rangle\cr
|4\rangle=\left|\left(\bar3,\bar3\right);(1,1)1\right\rangle\quad&
|8\rangle=\left|\left(\bar3,6\right);(1,0)1\right\rangle\cr}
\end{equation}
For $({\overline Q}ssud)$, we use the labeling $1=Q$, $2=s$, and $3=q_0$ to
list the matrix elements
\begin{eqnarray}
\left(\delta E_{\rm e}\right)_{1,1}&=&
{8\alpha_{\rm
c}\over3R}\left(f_{1,1}+f_{2,2}+f_{3,3}-f_{1,2}-f_{1,3}-f_{2,3}\right)\nonumber\\
\left(\delta E_{\rm e}\right)_{2,2}&=&
{4\alpha_{\rm
c}\over3R}\left(2f_{1,1}+2f_{2,2}+5f_{3,3}+f_{1,2}-5f_{1,3}-5f_{2,3}\right)\nonumber\\
\left(\delta E_{\rm e}\right)_{3,3}&=&
{4\alpha_{\rm
c}\over3R}\left(2f_{1,1}+5f_{2,2}+2f_{3,3}-5f_{1,2}+f_{1,3}-5f_{2,3}\right)\nonumber\\
\left(\delta E_{\rm e}\right)_{4,4}&=&\left(\delta E_{\rm
e}\right)_{7,7}=\left(\delta E_{\rm e}\right)_{1,1}\\
\left(\delta E_{\rm e}\right)_{5,5}&=&\left(\delta E_{\rm
e}\right)_{8,8}=\left(\delta E_{\rm e}\right)_{2,2}\nonumber\\
\left(\delta E_{\rm e}\right)_{6,6}&=&\left(\delta E_{\rm
e}\right)_{3,3}\nonumber
\end{eqnarray}
and
\begin{eqnarray}
\left(\delta E_{\rm m}\right)_{1,1}&=&
{8\alpha_{\rm c}\over R}\left(g_{2,2}+g_{3,3}-2g_{2,3}\right)\nonumber\\
\left(\delta E_{\rm m}\right)_{1,4}&=&
-{8\sqrt2\,\alpha_{\rm c}\over R}\left( g_{1,2}-g_{1,3}\right)\nonumber\\
\left(\delta E_{\rm m}\right)_{1,6}&=&
{12\sqrt2\,\alpha_{\rm c}\over R}g_{1,2}\nonumber\\
\left(\delta E_{\rm m}\right)_{1,8}&=&
-{12\sqrt2\,\alpha_{\rm c}\over R}g_{1,3}\nonumber\\
\left(\delta E_{\rm m}\right)_{2,2}&=&
{4\alpha_{\rm c}\over R}\left(2g_{2,2}-g_{3,3}-10g_{2,3}\right)\nonumber\\
\left(\delta E_{\rm m}\right)_{2,3}&=&
-{12\sqrt3\,\alpha_{\rm c}\over R}g_{2,3}\nonumber\\
\left(\delta E_{\rm m}\right)_{2,5}&=&
{4\sqrt2\,\alpha_{\rm c}\over R}\left(g_{1,2}+5g_{1,3}\right)\nonumber\\
\left(\delta E_{\rm m}\right)_{2,7}&=&
-{12\sqrt2\,\alpha_{\rm c}\over R}g_{1,3}\nonumber\\
\left(\delta E_{\rm m}\right)_{3,3}&=&
{12\alpha_{\rm c}\over R}\left(g_{2,2}-2g_{3,3}\right)\\
\left(\delta E_{\rm m}\right)_{3,7}&=&
-{12\sqrt6\,\alpha_{\rm c}\over R} g_{1,2}\nonumber\\
\left(\delta E_{\rm m}\right)_{4,4}&=&
{8\alpha_{\rm c}\over R}\left(
g_{2,2}+g_{3,3}-g_{1,2}-g_{1,3}-g_{2,3}\right)\nonumber\\
\left(\delta E_{\rm m}\right)_{4,6}&=&
{24 \alpha_{\rm c}\over R}\left(g_{1,2}+g_{2,3}\right)\nonumber\\
\left(\delta E_{\rm m}\right)_{4,8}&=&
{24 \alpha_{\rm c}\over R}\left(g_{1,3}+ g_{2,3}\right)\nonumber\\
\left(\delta E_{\rm m}\right)_{5,5}&=&
 {4\alpha_{\rm c}\over R}\left(
2g_{2,2}-g_{3,3}+g_{1,2}-5g_{1,3}-5g_{2,3}\right)\nonumber\\
\left(\delta E_{\rm m}\right)_{5,7}&=&
{24 \alpha_{\rm c}\over R}\left(g_{1,3}+ g_{2,3}\right)\nonumber\\
\left(\delta E_{\rm m}\right)_{6,6}&=&
{4 \alpha_{\rm c}\over R}\left(3g_{2,2}+2g_{3,3}+2g_{1,3}\right)\nonumber\\
\left(\delta E_{\rm m}\right)_{6,8}&=&
{12\alpha_{\rm c}\over R} g_{2,3} \nonumber\\
\left(\delta E_{\rm m}\right)_{7,7}&=&
 {8\alpha_{\rm c}\over R}\left( g_{2,2}-3g_{3,3}-2g_{1,2}\right)\nonumber\\
\left(\delta E_{\rm m}\right)_{8,8}&=&
 {4\alpha_{\rm c}\over R}\left( 2g_{2,2}+3g_{3,3}+2g_{1,2}\right).\nonumber
\end{eqnarray}
For $({\overline Q}uuds)$, we use  the labeling $1=Q$, $2=q_0$, and $3=s$ to
list the matrix elements \begin{eqnarray}
\left(\delta E_{\rm e}\right)_{1,1}&=&
{4\alpha_{\rm
c}\over3R}\left(2f_{1,1}+3f_{2,2}+2f_{3,3}-3f_{1,2}-f_{1,3}-3f_{2,3}\right)\nonumber\\
\left(\delta E_{\rm e}\right)_{1,2}&=&
{2\sqrt2\,\alpha_{\rm
c}\over3R}\left(-f_{2,2}+f_{1,2}-f_{1,3}+f_{2,3}\right)\nonumber\\
\left(\delta E_{\rm e}\right)_{2,2}&=&
{2\alpha_{\rm
c}\over3R}\left(4f_{1,1}+3f_{2,2}+4f_{3,3}-3f_{1,2}-5f_{1,3}-3f_{2,3}\right)\nonumber\\
\left(\delta E_{\rm e}\right)_{3,3}&=&
{2\alpha_{\rm
c}\over3R}\left(4f_{1,1}+9f_{2,2}+4f_{3,3}-9f_{1,2}+f_{1,3}-9f_{2,3}\right)\nonumber\\
\left(\delta E_{\rm e}\right)_{4,4}&=&\left(\delta E_{\rm
e}\right)_{7,7}=\left(\delta E_{\rm e}\right)_{1,1}\\
\left(\delta E_{\rm e}\right)_{5,5}&=&\left(\delta E_{\rm
e}\right)_{8,8}=\left(\delta E_{\rm e}\right)_{2,2}\nonumber\\
\left(\delta E_{\rm e}\right)_{6,6}&=&\left(\delta E_{\rm
e}\right)_{3,3}\nonumber\\
\left(\delta E_{\rm e}\right)_{4,5}&=&\left(\delta E_{\rm
e}\right)_{7,8}=\left(\delta E_{\rm e}\right)_{1,2}\nonumber
\end{eqnarray}
and
\begin{eqnarray}
\left(\delta E_{\rm m}\right)_{1,2}&=&
-{12\sqrt2\,\alpha_{\rm c}\over R}\left( g_{2,2}-g_{2,3}\right)\nonumber\\
\left(\delta E_{\rm m}\right)_{1,3}&=&
  {6\sqrt6\,\alpha_{\rm c}\over R}\left(  g_{2,2}-g_{2,3}\right)\nonumber\\
\left(\delta E_{\rm m}\right)_{1,4}&=&
-{4\sqrt2\,\alpha_{\rm c}\over R}\left( g_{1,2}-g_{1,3}\right)\nonumber\\
\left(\delta E_{\rm m}\right)_{1,5}&=&
-{12 \alpha_{\rm c}\over R}\left( g_{1,2}-g_{1,3}\right)\nonumber\\
\left(\delta E_{\rm m}\right)_{1,6}&=&
{12\sqrt2\,\alpha_{\rm c}\over R}g_{1,2}\nonumber\\
\left(\delta E_{\rm m}\right)_{1,7}&=&
 {4\alpha_{\rm c}\over R}\left( g_{1,2}-g_{1,3}\right)\nonumber\\
\left(\delta E_{\rm m}\right)_{1,8}&=&
-{6\sqrt2\,\alpha_{\rm c}\over R}\left( g_{1,2}+g_{1,3}\right)\nonumber\\
\left(\delta E_{\rm m}\right)_{2,2}&=&
-{12\alpha_{\rm c}\over R}\left( g_{2,2}+2g_{2,3}\right)\nonumber\\
\left(\delta E_{\rm m}\right)_{2,3}&=&
-{6\sqrt3\,\alpha_{\rm c}\over R}\left( g_{2,2}+g_{2,3}\right)\nonumber\\
\left(\delta E_{\rm m}\right)_{2,4}&=&
-{12\alpha_{\rm c}\over R}\left( g_{1,2}-g_{1,3}\right)\nonumber\\
\left(\delta E_{\rm m}\right)_{2,5}&=&
{2\sqrt2\,\alpha_{\rm c}\over R}\left(7g_{1,2}+5g_{1,3}\right)\nonumber\\
\left(\delta E_{\rm m}\right)_{2,7}&=&
-{6\sqrt2\,\alpha_{\rm c}\over R}\left(g_{1,2}+g_{1,3}\right)\nonumber\\
\left(\delta E_{\rm m}\right)_{2,8}&=&
{10\alpha_{\rm c}\over R}\left(g_{1,2}-g_{1,3}\right)\nonumber\\
\left(\delta E_{\rm m}\right)_{3,3}&=&
{12\alpha_{\rm c}\over R}\left(g_{2,2}-2g_{3,3}\right)\nonumber\\
\left(\delta E_{\rm m}\right)_{3,6}&=&
{2\sqrt3\,\alpha_{\rm c}\over R}\left(g_{1,2}- g_{1,3}\right)\nonumber\\
\left(\delta E_{\rm m}\right)_{3,7}&=&
-{12\sqrt6\,\alpha_{\rm c}\over R} g_{1,2}\\
\left(\delta E_{\rm m}\right)_{4,4}&=&
{4\alpha_{\rm c}\over
R}\left(g_{2,2}-3g_{1,2}-g_{1,3}+g_{2,3}\right)\nonumber\\
\left(\delta E_{\rm m}\right)_{4,5}&=&
-{6\sqrt2\,\alpha_{\rm c}\over R}\left(
g_{2,2}-g_{1,2}+g_{1,3}-g_{2,3}\right)\nonumber\\
\left(\delta E_{\rm m}\right)_{4,6}&=&
{12\alpha_{\rm c}\over
R}\left(g_{2,2}+2g_{1,2}+g_{2,3}\right)\nonumber\\\left(\delta E_{\rm
m}\right)_{4,7}&=&
{4\sqrt2\,\alpha_{\rm c}\over R}\left(
g_{2,2}-g_{1,2}+g_{1,3}-g_{2,3}\right)\nonumber\\
\left(\delta E_{\rm m}\right)_{4,8}&=&
{12\alpha_{\rm c}\over R}\left(g_{2,2}+g_{1,2}+g_{1,3}
+g_{2,3}\right)\nonumber\\
\left(\delta E_{\rm m}\right)_{5,5}&=&
-{2\alpha_{\rm c}\over R}\left(
g_{2,2}+3g_{1,2}+5g_{1,3}+7g_{2,3}\right)\nonumber\\
\left(\delta E_{\rm m}\right)_{5,6}&=&
-{6\sqrt2\,\alpha_{\rm c}\over R}\left(g_{2,2}- g_{2,3}\right)\nonumber\\
\left(\delta E_{\rm m}\right)_{5,7}&=&
{12\alpha_{\rm c}\over R}\left(g_{2,2}+g_{1,2}+g_{1,3}+
g_{2,3}\right)\nonumber\\
\left(\delta E_{\rm m}\right)_{5,8}&=&
{10\sqrt2\,\alpha_{\rm c}\over
R}\left(g_{2,2}-g_{1,2}+g_{1,3}-g_{2,3}\right)\nonumber\\
\left(\delta E_{\rm m}\right)_{6,6}&=&
{4 \alpha_{\rm c}\over R}\left(3g_{2,2}+ g_{1,2}+
g_{1,3}+2g_{2,3}\right)\nonumber\\
\left(\delta E_{\rm m}\right)_{6,7}&=&
-{6\sqrt2\,\alpha_{\rm c}\over R}\left(g_{2,2}-g_{2,3}\right)\nonumber\\
\left(\delta E_{\rm m}\right)_{6,8}&=&
{6\alpha_{\rm c}\over R} \left(g_{2,2}+g_{2,3}\right) \nonumber\\
\left(\delta E_{\rm m}\right)_{7,7}&=&
{8\alpha_{\rm c}\over R}\left(g_{2,2}-2g_{1,2}-3g_{2,3}\right)\nonumber\\
\left(\delta E_{\rm m}\right)_{8,8}&=&
 {4\alpha_{\rm c}\over R}\left( 2g_{2,2}+2g_{1,2}+3g_{2,3}\right).\nonumber
\end{eqnarray}

\begin{table}
\caption{Some ordinary or charmed hadrons  in the simple bag model of
Sec.~{\protect\ref{Bag-Heavy-Light}}.}\par
\vskip .2cm
\begin{tabular}{ccc}
State&Mass&Exp.\\
\tableline
$\pi(q_0\bar q_0)$&0.11&0.14\\
$\rho(q_0\bar q_0$&0.77&0.77\\
$K(q_0\bar s)$&0.51&0.50\\
$K^*(q_0\bar s)$&0.92&0.89\\
$\Phi(s\bar s)$&1.07&1.02\\
$N(q_0 q_0q_0)$&0.95&0.94\\
$\Delta(q_0 q_0q_0)$&1.23&1.23\\
$\Lambda(sq_0q_0)$&1.12&1.11\\
$\Sigma(sq_0q_0)$&1.16&1.19\\
$\Sigma^*(sq_0q_0)$&1.34&1.38\\
$\Xi(ssq_0)$&1.31&1.32\\
$\Xi^*(ssq_0)$&1.49&1.53\\
$\Omega^-(sss)$&1.68&1.67\\
$D(c\bar q_0)$& 1.85&1.87\\
$D^*(c\bar q_0)$& 2.03&2.01\\
$D_s(c\bar s)$& 1.95&1.97\\
$D_s^*(c\bar s)$& 2.11&2.11\\
$\Lambda_c(cq_0q_0)$& 2.30&2.28\\
$\Sigma_c(cq_0q_0)$& 2.40&2.45\\
$\Xi_c(csq_0)$& 2.46&2.47\\
\end{tabular}
\label{Tab0}
\end{table}
\begin{table}
\caption{Various contributions to the energy of $(cs\bar u\bar u)$, $(c\bar
u)$, and $(s\bar u)$: volume energy, quark mass and kinetic energy, zero-point
energy, electric and magnetic terms, and centre-of-mass corrections,
respectively, in GeV. Also shown are the bag radius, in GeV$^{-1}$, and the
coupling $\alpha_{\rm c}$.}\par\vskip .2cm
\begin{tabular}{ccccccccc}
State&radius&$\alpha_{\rm c}$&volume&quark&zero-point&el.+mag.&c.o.m.&mass
\\
\tableline
$s\bar u$& 4.15&0.37&0.11&1.13&$-0.14$&$-0.21$&$-0.39$&0.51\\
$c\bar u$& 4.23&0.38&0.12&0.48&$-0.14$&$-0.61$&&1.85\\
$cs\bar u\bar u$& 5.85&0.49&0.31&1.21&$-0.10$&$-0.63$&&2.79\\
\end{tabular}
\label{Tab1}
\end{table}
\begin{table}
\caption{Four-quark configurations with charm. The mass (in GeV) is compared to
the lowest dissociation threshold.}\par\vskip .2cm
\begin{tabular}{cccc}
\multicolumn{2}{c}{State}&\multicolumn{2}{c}{Threshold}\\
Content&Mass&Content&Mass\\
\tableline
$cs\bar s\bar s$& 3.07&$(c\bar s)+(s\bar s)$&2.69\\
$cq_0(\bar q_0\bar q_0)_{I=1}$& 2.61&$(c\bar q_0)+(q_0\bar q_0)$&1.96\\
$cq_0\bar s\bar s$& 2.90&$(c\bar s)+(q_0\bar s)$&2.46\\
$cs(\bar q_0\bar q_0)_{I=1}$& 2.79&$(c\bar q_0)+(s\bar q_0)$&2.36\\
$cq_0(\bar q_0\bar q_0)_{I=0}$& 2.36&$(c\bar q_0)+(q_0\bar q_0)$&1.96\\
$cs(\bar q_0\bar q_0)_{I=0}$& 2.59&$(c\bar q_0)+(s\bar q_0)$&2.36\\
$cq_0(\bar s\bar q_0)$& 2.51&$(c\bar s)+(q_0\bar q_0)$&2.06\\
$cs(\bar q_0\bar s)$& 2.73&$(c\bar s)+(s\bar q_0)$&2.46\\
 \end{tabular}
\label{Tab2}
\end{table}
\begin{table}
\caption{Five-quark configurations with charm or beauty. The mass (in GeV) is
compared to the lowest dissociation threshold.}\par\vskip .2cm
\begin{tabular}{cccc}
\multicolumn{2}{c}{State}&\multicolumn{2}{c}{Threshold}\\
Content&Mass&Content&Mass\\
\tableline
$\bar cssud$& 3.16&$(\bar cs)+(sud)$&3.07\\
$\bar cuuds$& 2.97&$(\bar cs)+(uud)$&2.89\\
$\bar bssud$& 6.58&$(\bar bs)+(sud)$&6.50\\
$\bar buuds$& 6.39&$(\bar bs)+(uud)$&6.32\\
\end{tabular}
\label{Tab3}
\end{table}
\begin{table}
\caption{Four-quark configurations with charm, and
the corresponding thresholds,
in a model where the zero-point energy parameter $Z_0$ is set to
zero.}\par\vskip .2cm \begin{tabular}{cccc}
\multicolumn{2}{c}{State}&\multicolumn{2}{c}{Threshold}\\
Content&Mass&Content&Mass\\
\tableline
$cs\bar s\bar s$& 3.17&$(c\bar s)+(s\bar s)$&3.04\\
$cq_0(\bar q_0\bar q_0)_{I=1}$& 2.71&$(c\bar q_0)+(q_0\bar q_0)$&2.49\\
$cq_0\bar s\bar s$& 3.00&$(c\bar s)+(q_0\bar s)$&2.83\\
$cs(\bar q_0\bar q_0)_{I=1}$& 2.89&$(c\bar q_0)+(s\bar q_0)$&2.74\\
$cq_0(\bar q_0\bar q_0)_{I=0}$& 2.46&$(c\bar q_0)+(q_0\bar q_0)$&2.49\\
$cs(\bar q_0\bar q_0)_{I=0}$& 2.69&$(c\bar q_0)+(s\bar q_0)$&2.74\\
$cq_0(\bar s\bar q_0)$& 2.62&$(c\bar s)+(q_0\bar q_0)$&2.58\\
$cs(\bar q_0\bar s)$& 2.83&$(c\bar s)+(s\bar q_0)$&2.83\\
 \end{tabular}
\label{Tab4}
\end{table}


\end {document}